\documentstyle[preprint,aps,prl, psfig]{revtex}
\begin{document} 
% the following two lines are to get a wide abstract
%@twocolumnfalse\endcsname

\title{Interaction of Reaction-Diffusion Fronts and Marangoni Flow on the Interface of Deep Fluid}

\author{L. M.~Pismen,\\
{\it Department of Chemical Engineering,}\\
{\it Technion -- Israel Institute of Technology, Technion City, Haifa 32000, Israel}}
% \date{}
\maketitle
\begin{abstract}

We consider a bistable reaction-diffusion system on the interface 
of deep fluid interacting with Marangoni flow. The method of 
matched asymptotic expansions is used to resolve the singularity at a 
sharp interface between the alternative states, and to compute the 
self-induced flow velocity advecting the domain boundary. 
It is shown that Marangoni flow 
serves as an efficient mechanism preventing the spread of the state with
a higher surface tension when it is dynamically favorable.
\end{abstract}


\vspace{-0.3in}
\vskip2pc]
\narrowtext

%\paragraph{Introduction.}

It has been long realized that chemical instabilities 
can both induce and be strongly affected
by hydrodynamic flows. In early experiments
 with BZ and other oscillatory reactions this
interaction led to a transition from propagating
 waves in confined systems to cellular
structures in systems with a free interface where
 Marangoni flow was excited by concenration
gradients of chemical origin \cite{bz}. Strong 
interaction between chemical and interfacial
instabilities has been the subject of a number 
of experimental studies \cite{expm}. A related 
direction was the study of chemical fronts 
affected by buoyancy \cite{expg}; both effects 
were apparently relevant in observations of patterns 
of photochemical origin \cite{avnir}. 

Following the linear analysis of chemical instability 
coupled with Marangoni flow
\cite{sanf}, theoretical analysis of the nonlinear development of 
reaction-convection instabilities has been restricted so far 
either to very shallow layers enabling 
lubrication approximation \cite{dagan}, 
or to systems close to a bifurcation point \cite{brand}; 
both approaches are inapplicable to the 
most practical case of sharp reaction fronts. 
The only other alternative was numerical modeling \cite{comp}. 

The most powerful technique for constructing and analysing
 non-equilibrium patterns of chemical origin is based on tracing 
the motion of sharp interfaces between alternative states of a 
bistable system interacting with a long-range field \cite{chph,lp}. 
Marangoni flow due to concentration gradients in the transitional 
region separating the two alternative states and, in its turn, advecting 
this boundary, well qualifies as a long-range controlling agent; 
its action is, however, highly nonlocal, which poses 
serious analytical difficulties. 

Recently, boundary integral technique was applied to computation 
of the motion of interfacial domains coupled to creeping motion of the 
underlying fluid \cite{gold,thess}. The effect of a sharp interface was, 
however, the emergence of a finite time singularity \cite{thess} that ruled 
out formation of stable structures. I shall show in this
Letter that the singularity can be indeed resolved when finite, though 
low diffusivity is taken into account. The method 
of matched asymptotic expansions (akin to that used for the analysis 
of motion of vortex lines \cite{piru}) will be applied to resolve the structure of the 
domain boundary, and to obtain a finite stationary velocity of the induced 
flow. I shall further show that inward flow can stabilize a solitary spot 
of a dynamically prevalent state with a higher surface tension. 

%\paragraph{Flow pattern.}

Consider creeping motion of a viscous incompressible fluid in an infinitely deep 
and infinitely extended
layer $z<0$ induced by the Marangoni force due to a given stationary
distribution of surface tension $\sigma({\bf x})$ on the undeformable free
boundary $z=0$. Since the vertical vorticity vanishes, the 2d horizontal
velocity vector {\bf v} and the vertical velocity $w$ are expressed,
respectively, as ${\bf v}=\nabla \chi_z,\; w=-\nabla^2 \chi$, where
$\nabla$ is the 2d vector differential operator. The velocity potential
$\chi({\bf x},z)$ is determined by solving the Stokes equation
\begin{equation}
(\nabla^2+\partial^2_z)^2\chi=0
\label{stokeseq}  \end{equation}
with the Marangoni boundary conditions at the free surface $z=0$:
\begin{equation}
\chi({\bf x},0) = 0, \;\;\;  \mu\chi_{zz}({\bf x},0)= -\nabla \sigma,
\label{stokes}  \end{equation}
where $\mu$ is dynamic viscosity.

The solution is found most readily by Fourier transform \cite{pozr}, yielding 
\begin{equation}
\hat\chi(k)= -\frac{1}{2k\mu}z e^{kz}\hat \sigma(k),
\label{chi}  \end{equation}
where hats denote Fourier transforms of the respective functions, and 
$k=|{\bf k}|$. The only quantity relevant for our purpose is the interfacial
velocity, which is computed as ${\bf v}=\nabla\phi$, where the interfacial 
flow potential $\phi({\bf x})= \chi_z({\bf x},0)$ can be expressed in
the form
\begin{equation}
\phi({\bf x})= \mu^{-1} \int G(|{\bf x}-\xi|) \sigma(\xi) d\xi,
\label{phi}  \end{equation}
where the integration is carried out over the entire free surface. The
kernel $G(r)$ depends only on the 2d distance $r=|{\bf x}-\xi|$, and is
obtained as the inverse transform of $\hat\chi_z(0)$:  
\begin{equation}
G(r) =\frac{1}{8\pi^2} \int k^{-1} e^{-i{\bf k}\cdot {\bf x}}  d^2{\bf x} = 
    \frac{1}{4\pi}  \int_0^\infty J_0(kr) dk =  \frac{1}{4\pi r}.
\label{g}  \end{equation}
It is convenient to express the velocity through the {\it gradient} of
surface tension:  
\begin{equation}
{\bf v}({\bf x}) = \mu^{-1} \int G({\bf x}-\xi) \nabla \sigma(\xi) d\xi.
\label{v}  \end{equation}
The solution exists provided $\sigma({\bf x})$ = const at $|{\bf x}| \to
\infty$. 

%\paragraph{Surface concentration.}

The distribution of the surface concentration $\theta$ of an insoluble 
surfactant on the free surface obeys the convective reaction-diffusion
equation, which we shall write in a dimensionless form 
\begin{equation}
\theta_t  =\mbox{$\frac{1}{2}$} \nabla^2 \theta - 
       \nabla\cdot({\bf c}\theta) + f(\theta).
\label{theta}  \end{equation}
Here $f(\theta)$ is the dimensionless net surfactant source due to chemical 
reactions and exchange with the gas phase; the time scale is the
characteristic reaction time $\tau$; the coordinates are scaled by the
diffusional length $\sqrt{2D\tau}$, where $D$ is the surface diffusivity;
${\bf c}={\bf v}\sqrt{\tau/2D}$ is the dimensionless velocity. We are
interested in the case when at least two stable stationary solutions exist;
therefore the function $f(\theta)$ should have three zeroes, and its
derivative at the smallest and largest zero should be negative. The
simplest function possessing this property is a cubic. By shifting and
rescaling, the two stable zeroes can be placed at $\theta=0$ and 
$\theta=1$. Thus, we can assume $f(\theta)=-\theta(\theta-q)(\theta-1)$,
where $0<q<1$.

If ${\bf c}=$ const, Eq.~(\ref{theta}) with the cubic source has a 
simple stationary solution
\begin{equation}
\theta=\frac{1}{1+e^{-x} }\equiv \frac{1}{2}\left(1+
\tanh\frac{x}{2}\right),    
\label{ths}  \end{equation}
corresponding to the velocity $c=-c_0=\frac{1}{2}-q$
 directed along the $x$ 
axis. The front translates normally to itself towards
 the upper state with the ``chemical'' 
speed $c_0$ in the coordinate frame comoving 
with the local interface flow velocity {\bf c}. 
Hence, $ c_0-{\bf c} \cdot {\bf n}$ is the displacement
 speed of a point on the front along 
the normal ${\bf n}$. This speed can be corrected 
to account  for the curvature effect. The
front is stationary when the interface is advected 
in the direction opposite to chemical propagation with the same 
speed. 

In the presence of Marangoni flow, Eq.~(\ref{theta}) becomes an 
integro-differential equation, where {\bf c} is expressed with the help of
Eq.~(\ref{v}) as  
\begin{equation}
{\bf c}({\bf x}) = - M \int G({\bf x}-\xi) \nabla \theta(\xi) d\xi,
\label{vc}  \end{equation}
where $M=(\Delta\sigma /\mu)\sqrt{\tau/2D})$ is the Marangoni number
based  on the characteristic diffusional length; $\Delta\sigma$ is
the difference of surface tension between the lower ($\theta=0$) and upper
($\theta=1$) stationary states (it is assumed that the surface tension
decreases linearly with growing concentration). 

If the front motion is sufficiently slow, so that the 
flow field remains quasistationary, 
Eq.~(\ref{vc}) applies, 
which gives a closed equation of the front displacement. 
This equation is {\em nonlocal} 
so that the motion is dependent on the instantaneous 
distribution of domains occupied by 
the alternative states in the entire region. 

%\paragraph{The singular limit.}

The thickness of the front region where the change of the surfactant 
concentration occurs is determined by the characteristic diffusional length
$\sqrt{2D/ \tau}$. On distances far exceeding this scale, the
concentration distribution can be considered as stepwise. 
Using the discontinuous function 
$\theta({\bf x})$ that assumes the two alternative 
values $\theta=0$ and $\theta=1$ in the 
domains separated by a boundary $\Gamma$ 
(presumed smooth but not necessarily 
simply connected) brings Eq.~(\ref{vc}) to the form 
\begin{equation}
{\bf c}({\bf x}) = - \frac{M}{4\pi} \oint_\Gamma  \frac{{\bf n}}{r(s)}  ds,
\label{vc0}  \end{equation}
where the contour is parametrized by the arc length 
$s$, and $r(s)=|{\bf x}-\xi(s)|$ is the 
distance from the reference point to a point on 
the contour; by convention, the normal is 
directed towards the domain occupied by the upper state. 

Eq.~(\ref{vc0}) is quite adequate for computing
 the flow velocity far from the fronts but 
cannot be applied to the problem of front dynamics,
 since the integral diverges on the front 
location. To compute the velocity induced within the
 front region, one has to remove a 
short arc segment from the integral Eq.~(\ref{vc0}),
 and apply on the rest of the contour Eq.~(\ref{vc}) 
containing a smooth concentration profile. On the removed 
segment, the finite diffusional length has taken into account, 
and both parts of the contour matched as in Ref \cite{piru}.

\begin{figure}
\psfig{figure=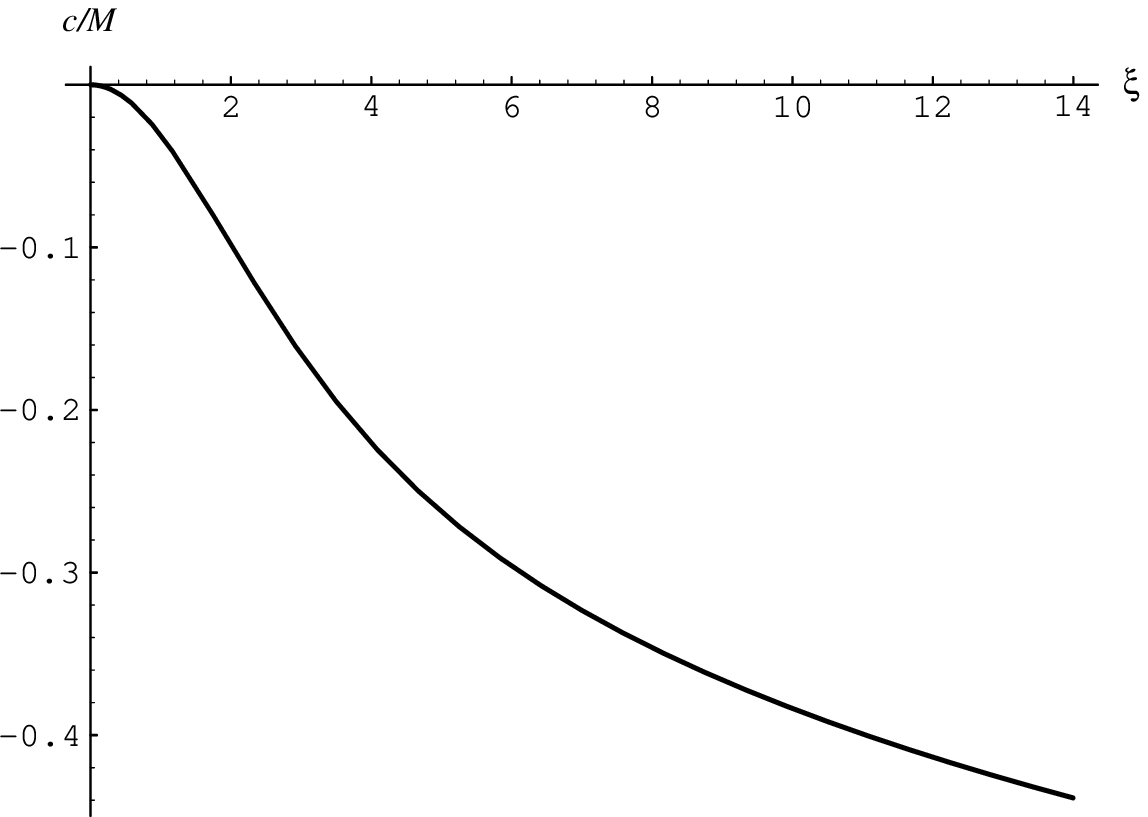}
\caption{The velocity profile in the vicinity of the front (the maximum
velocity is taken as the zero level).}
\label{fig1}
\end{figure}

We presume that the local curvature radius of the 
front is of the same order of magnitude 
as the characteristic macroscopic scale $L$, and far 
exceeds the diffusional length. Then the concentration gradient
 is directed normally to the 
front, and the contour segments are almost rectilinear on
 distances $s \ll L$. In 
accordance to a common procedure of the theory of front
 dynamics \cite{chph,lp}, we 
transform to a local coordinate frame comoving with the
 front, and take a certain contour, say,
$\theta=q$ as the origin of the normal coordinate $\xi$. 
The origin of the tangential coordinate $\eta$ is taken at the foot of
 the normal drawn from the reference 
point. Consider a point with a normal coordinate 
$\xi \ll L$, and cut from the contour 
integral in Eq.~(\ref{vc0}) a small arc of the length 
$2\delta$, where  $\xi \ll \delta \ll L$. 
On the remaining  part of the contour, $\xi$ 
can be neglected; thus, the velocity induced at 
the reference point by the ``far'' contour is 
determined by Eq.~(\ref{vc0}). The normal 
velocity computed in this way diverges 
logarithmically at $\delta \to 0$, and can be expressed as 
\begin{equation}
c^{(o)} \asymp - \frac{M}{2\pi}\ln \frac{C}{\delta}, 
\label{c2o}  \end{equation}
where $C$ is a constant dependent on the 
shape of the contour in the far region. In the near 
region, we introduce a stretched coordinate 
$\eta =s/\xi$, and compute the inner integral as  
\begin{eqnarray}
c^{(i)}(\xi) &=&  -\frac{M}{2\pi}  \lim_{\delta/\xi \to \infty}
    \int_{-\infty}^{\infty} \theta'(\xi-\zeta) d\zeta 
     \int_0^{\delta /\zeta}\frac{d\eta}{\sqrt{\zeta^2 +\eta^2}} \nonumber \\ 
    &=&  -\frac{M}{2\pi} \left[ \ln (2\delta) - 
     \int_{-\infty}^{\infty} \ln |\zeta|\theta'(\xi-\zeta)  d\zeta \right].
\label{c2i}  \end{eqnarray}
The auxilliary value $\delta$ falls out 
when both integrals are added up.
The resulting finite expression for 
the flow velocity in the front region is
\begin{equation}
c(\xi) =  -\frac{M}{2\pi} \ln \frac{C}{\beta(\xi) }; \;\;\; 
    \ln \beta(\xi) = \int_{-\infty}^{\infty}
 \ln |\xi -\zeta|  \theta'(\zeta)  d\zeta.
\label{c20}  \end{equation}
Since $ \theta'(\zeta) \to 0$ at $\zeta \to\infty$, 
$\beta(\xi) \asymp \xi$ at $\xi \gg 1$.
The function $ \beta(\xi)$  computed numerically using the analytical
solution  $\theta'(\xi)=\frac{1}{4}
 \mbox{sech}^2\frac{\xi }{2}$ corresponding 
to Eq.~(\ref{ths}) is plotted in Fig.~\ref{fig1}. 

%\paragraph{A solitary spot.}

\begin{figure}
\psfig{figure=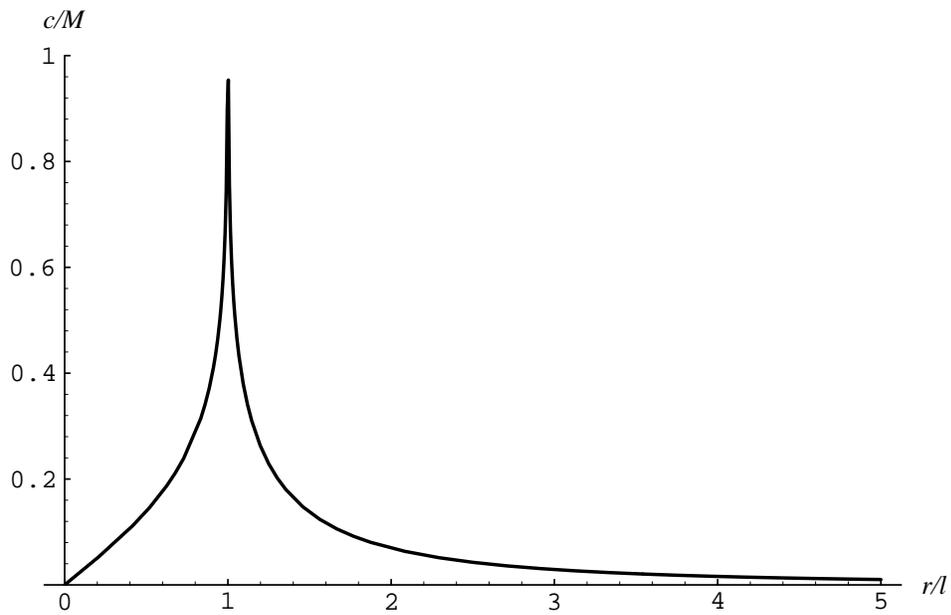}
\caption{The interfacial radial velocity in and around a solitary spot.}
\label{fig2}
\end{figure}

In view of the nonlocal character of the induced 
motion, only a symmetric configuration 
may be stationary. Consider a spot of the radius 
$L$ occupied by the lower state on the infinite
 interface occupied by the upper state. We presume
 that the spot radius is large when
measured on the diffusional scale, so that 
the dimensionless radius
$l=L/\sqrt{2D\tau} \gg 1$. The induced surface 
flow velocity is computed using Eq.~(\ref{vc0}) as
\begin{eqnarray}
c(\rho)&=& -\frac{M}{2\pi} \int_0^\pi \frac{\cos\alpha d\alpha}
   {\sqrt{1 +(\rho/l)^2-2(\rho/l)\cos\alpha}}
\nonumber \\
&=& \left\{ \begin{array}{lll} 
   - \frac{M}{\pi} \left[{\bf K}\left(\frac{l}{\rho}\right)  -
    {\bf E}\left(\frac{l}{\rho}\right) \right]  & \mbox{at} & \rho>l,\\ 
& & \\
   - \frac{Ml}{\pi \rho} \left[{\bf K}\left(\frac{\rho}{l}\right) -
    {\bf E}\left(\frac{\rho}{l}\right) \right]  & \mbox{at} & \rho<l . 
\end{array} \right.
\label{c2}  \end{eqnarray}
where $\rho, \alpha$ are polar coordinates, and {\bf K}, {\bf E} are
complete elliptic integrals.  The radial flow velocity vanishes at $r=0$ and
decreases $\propto 1/r$ at $r\to\infty$ (Fig.~\ref{fig2}). 

As expected, there is a divergence in the vicinity of the front: 
\begin{equation}
c \asymp -\frac{M}{2\pi} \ln \left|\frac{8l}{e^2\xi}\right| \;\;
  \mbox{at }\; \xi\to 0 , 
\label{c2a}  \end{equation}
where $\xi =\rho -l$, and $l\gg \xi = O(1)$. 
The divergence is resolved using the 
asymptotic procedure described above.
The resulting finite expression for the flow 
velocity  at distances $\xi=O(1)$ from the spot 
circumference coincides with Eq.~(\ref{c20}) where $C=8l/e^2$. 
In the outer limit $x\to\infty$, Eq.~(\ref{c20})
coincides with the inner limit (\ref{c2a}) of the outer solution. 
The uniformly valid composite expansion can be obtained by 
adding up Eqs.~(\ref{c20}), (\ref{c2}) and
 extracting their common limit (\ref{c2a}):  
\begin{equation}
c(\rho)= \left\{ \begin{array}{lll} 
   - \frac{M}{\pi} \left[{\bf K}\left(\frac{l}{\rho}\right)  -
    {\bf E}\left(\frac{l}{\rho}\right) +
 \frac{1}{2} \ln \frac{\rho -l}{\beta(\rho -l)}\right]  
       & \mbox{at} & \rho>l,\\   & & \\
   - \frac{M}{\pi } \left[\frac{l}{\rho}{\bf K}\left(\frac{\rho}{l}\right) -
      \frac {l} {\rho}{\bf E}\left(\frac{\rho}{l}\right) +
     \frac{1}{2} \ln \frac{l- \rho }{\beta(l-\rho)}\right] 
  & \mbox{at} & \rho<l .    \end{array} \right.
\label{c2c}  \end{equation}

For the purpose of semi-quantitative estimation, one can assume that the 
front propagation is mostly affected by the maximum advection velocity in
the front region, which is observed at the location where the concentration
gradient is also at its maximum. Using the numerical value $\beta(0)=
0.882$, the maximum velocity is computed as $c_{m}=-(M/2\pi)\ln(0.955l)$.
Since the dependence on $l$ is logarithmical,
 this velocity may be of $O(1)$ on the 
diffusional scale, and balance the intrinsic front velocity
$c_0=q-\frac{1}{2}$ when $l \gg 1$ but $\ln l$ is not exceedingly
large. The radius of a stationary spot is computed then as  
\begin{equation}
L= 1.047 \sqrt{2D\tau} \exp  \frac{\pi (2q-1)}{M}. 
\label{l1}  \end{equation}
Since the variable flow velocity is actually smaller than its maximal
value, a somewhat larger radius might be actually necessary to induce
the flow counterbalancing the speed of the chemical front; thus, the above
expression gives a lower estimate of the spot size; the correction would
only amount, however, to modifying the numerical coefficient. The spot
should be stable to homogeneous perturbations, since an increase of the
radius would cause an increased inward flow restoring the stationary state.
The curvature correction to the intrinsic speed of the chemical front is
of $O(l^{-1})$, and can be neglected. I checked the stability using the method 
of boundary perturbations similar to that of Ref.~\cite{lp}. 
A reverse set-up, with a spot
occupied by the upper state (with a lower surface tension) is treated
formally in the same way, but, as the flow in this case is outward, the
stationary state is evidently unstable. A similar instability was observed in
experiment with radial spreading of a surfactant \cite{perm}.  

A  rational expansion can be carried out 
for the case when the Marangoni number
is small, and the two alternative states are close to the Maxwell
construction. The most interesting situation arises when front curvature
effects are also of the same order of magnitude. Thus, we can take $l^{-1}$
as the small parameter of the expansion, and set $q=\frac{1}{2} + q_1/l, \;
M= M_1/l$.  Expanding $\theta = \theta_0 + l^{-1} \theta_1 + \ldots, \; c=
l^{-1}  c_1 + \ldots$, yields in the zero order $\theta_0$ defined by
Eq.~(\ref{ths}). The stationary first order equation valid in the vicinity
of the boundary of a circular spot with the radius $l$ is
\begin{equation}
\mbox{$\frac{1}{2}$} \theta_1^{''}(\rho) + f'(\theta_0) \theta_1 + 
\mbox{$\frac{1}{2}$} \theta_0^{'}(\rho)  - 
  q_1 \theta_0(1-\theta_0) - (c_1 \theta_0)^{'} = 0 .
\label{theta1}  \end{equation}
The homogeneous part of this equation 
has a zero eigenvalue corresponding to
the translational symmetry of the zero-order equation; the corresponding
eigenfunction is $\theta_0^{'}(\rho)=\frac{1}{2}\mbox{sech}^2\frac{\xi }{2}; \;
\xi=\rho-l$. The solvability condition of Eq.~(\ref{theta1}) requires that its
inhomogeneous part be orthogonal to this eigenfunction. This yields a
relation between $q_1$ and $M_1$ that has to be satisfied at a stationary
front. Since the eigenfunction falls off sharply at $\xi \gg 1$, the
asymptotic expression for the velocity valid in the front region,
Eq.~(\ref{c20}) with $M$ replaced by $M_1$
 can be used for $c_1$. Computing
the integrals, we express the solvability condition as
\begin{equation}
 \frac{1}{2} - q_1 +  \frac{M_1}{2\pi} \ln \frac{l}{l_0}= 0,
\label{solv}  \end{equation}
 where
\begin{equation}
 \ln l_0 = 6 \int_{-\infty }^\infty 
 \theta_0^{''}(\xi) \theta_0(\xi)d\xi
\int_{-\infty }^\infty  \ln |\xi-\zeta| \theta_0^{'}(\zeta) d\zeta = 1.344.
\label{solvl0}  \end{equation}
The first term in Eq.~(\ref{solv}) describes the curvature effect. If it
is neglected, Eq.~(\ref{l1}) is recovered, but with a larger numerical
coefficient $\exp(l_0)=1,969$. 

If there is a number of spots removed at distances far exceeding the 
diffusional length, the induced flow is additive, since the Stokes problem 
is linear. Separate spots are attracted one to the other by the induced inward 
flow; clearly, the system shows a tendency to aggregation but coalescing
spots should srink again to the same stable radius at which the inward flow
is compensated by the chemical propagation speed. Thus, Marangoni flow 
serves as an efficient mechanism preventing the spread of the state with
a higher surface tension when it is dynamically favorable.

{\em Acknowledgement.} This research has been supported by the 
Fund for Promotion of Research at the Technion.

\end{document}